# Scheduling and Checkpointing optimization algorithm for Byzantine fault tolerance in Cloud Clusters

**Sathya Chinnathambi**
Department of Computing, Coimbatore Institute of Technology,
Coimbatore-14.Tamil Nadu, India,
**Dr Agilan Santhanam**
Department of Physics, Coimbatore Institute of Technology,
Coimbatore-14, Tamil Nadu, India.

**Abstract**

Cloud Computing distinguishes itself from other distributed computing paradigm through offering services on-demand basis without any geographical restrictions. This revolutionizes the computing by offering services to wide array of customers starting from casual user to highly business oriented Industries. In spite of its capabilities, Cloud Computing still struggle with handling wide array of faults, this causes loss of credibility to Cloud Computing. Among those faults Byzantine faults offers serious challenge to fault tolerance mechanism, because it often go undetected at the initial stage and it can easily propagate to other VMs before a detection is made. Consequently some of the mission critical application such as air traffic control, online baking etc still staying away from the cloud for such reasons. However if a Byzantine faults is not detected and tolerated at initial stage then applications such as big data analytics can go completely wrong in spite of hours of computations performed by the entire cloud. Therefore in the previous work a fool-proof Byzantine fault detection has been proposed, as a continuation this work designs a scheduling algorithm (WSSS) and checkpoint optimization algorithm (TCC) to tolerate and eliminate the Byzantine faults before it makes any impact. The WSSS algorithm keeps track of server performance which is part of Virtual Clusters to help allocate best performing server to mission critical application. WSSS therefore ranks the servers based on a counter which monitors every Virtual Nodes (VN) for time and performance failures. The TCC algorithm works to generalize the possible Byzantine error prone region through monitoring delay variation to start new VNs with previous checkpointing. Moreover it can stretch the state interval for performing and error free VNs in an effect to minimize the space, time and cost overheads caused by checkpointing. The analysis is performed with plotting state transition and CloudSim based simulation. The result shows TCC reduces fault tolerance overhead exponentially and the WSSS allots virtual resources effectively.

**Keywords:** Cloud Computing, Byzantine Faults, Checkpointing, Scheduling, Fault Tolerance.

1. Introduction

Cloud computing has revolutionized the distributed computing model with service on-demand and pay-as-you-go features [5][17]. These features facilitate the businesses with quick provisioning of unexpected peak demands through availing a resourceful and cheap cloud service [8][23]. Simplicity behind deploying cloud services makes it attractive for budding business to pioneering businesses. However, cloud services should be reliable to retain the Quality of

Service (QoS) requirements [4][20]. At functional level the cloud computing constitutes various Virtual Machines (VMs) [27] running across data centers that may be placed in diverse geographical locations to form a virtual cloud cluster which comprises multi-tenants [16]. Single or more such virtual cloud cluster can be formed dynamically to scale boundaries in accomplishing a single mission. This multi-tenancy model makes the cloud computing more prone to various security breaches than other distributed computing model [4][9][15]. Among those, the most challenging aspect is the attacker's capability to mask a breach as a Byzantine fault which most of the times causes serious damages to Cloud environment [1]. Since it induces itself evasively and can remain undetected. Moreover it can spread from one VM to another quickly. However other faults can be detected but byzantine faults remains elusive and causes serious damages, since the system keeps working even with the induced faults [21]. Moreover Byzantine fault can be used as the mode of propagation to cause VM, server, network, application and complete cloud failures [1]. However Cloud services are on-demand and paid for every unit time. Therefore the reliability assurance offered is high but due to failures such as Crash faults, Process Failure, Application failure, Network failure, Node failure, server failure etc [7][17] it becomes extremely difficult to maintain the required level of performance in fulfilling the QoS agreements. This could cause the customers discontent towards the cloud service providers (CSP).

However any fault handling mechanism in cloud can be categorized as Fault detection, Fault removal, fault prevention, fault forecasting, and fault tolerance [1]. Among those fault detection is the crucial and initial fault handling mechanism. However due to the evasive nature of the Byzantine faults it becomes tedious to detect it in the initial state. Therefore in the previous work a hash based delay sensitive byzantine fault checking mechanism has been proposed which is proved to be capable of detecting the byzantine faults effectively [1]. As the name implies other mechanism thrives to deal with the faults at various stages either in proactive or reactive fashion. Among those the fault prevention is vital since it tries to prevent the faults with proper defensive mechanism. However often the faults are developed to overcome the existing defense system therefore fault tolerance becomes crucial in the cloud systems as a contingency fault handling mechanism [26]. Apart from other techniques Fault tolerance is an attempt to ensure service continuity in spite of the fault occurrences. Fault tolerance is an attempt to enhance the cloud reliability from the practical implications [22]. There are various fault tolerance mechanisms such as checkpointing, replication, task Migration, Self- Healing, Safety-bag checks, Retry, Task Resubmission, Reconfiguration, Masking etc [6][7][13][22].

Among those in Cloud Services the Checkpointing is a widely adapted fault tolerance mechanism [20]. The checkpointing techniques frequently saves the state of every VM as image files and are readily deployable if any VM or cluster of VMs fails [2]. However to maintain collective checkpointing for an error prone situation where failure is expected, requires intense checkpointing for every minimized time intervals. The smaller time interval becomes more space and time consuming the checkpointing can be. Moreover dividable tasks such as in case of big data analytics [19]; if any one VM generates erroneous output it can be passed on to others and thus the entire output can be corrupted [14]. In such cases checkpointing is tedious and causes tremendous performance overhead. However there is only limited number of research publications were available for checkpointing improvisation, still the problem of strategizing the checkpointing remains an open challenge. Therefore checkpointing is done as an all inclusive task which means if a failure is expected the number of checkpointing is increased for every VMs invariably [29].

Large scale data processing such as big data analytics are often run on Cloud computing platforms due to vast processing requirements which cannot be achieved by stationary facility [11][14]. Such cases involve automatic distribution of input as many blocks often called jobs [3][4]. The jobs can be further sub divided as tasks which are made suitable for VM level processing [12]. However task is not always considered as the sub-division of jobs but in some cases it is seen as the representation of jobs. However the job is often fragmented into various tasks to suit the VMs. The sub division of Application to tasks or workloads can involve 'n' levels before it becomes process-able by VMs. In this scenario even if a single VM generate an error it can create a dominos effect by spreading to other VMs [5]. Moreover those are delay sensitive application because if a VM takes more time to complete a task then it can delay the complete application. Moreover due to uncertainty that revolves around byzantine error detection mostly checkpointing is rollbacked to few or initial stages and individual task migration is often replaced with complete job migration even if an error is detected at the individual VN.

However to overcome these problems a Scheduling algorithm to monitor the performance of the VNs online to rate the physical server has been proposed to identify the appropriate VNs for mission critical applications. Moreover effective checkpointing algorithm to sustain the task and job migration applications with minimal overhead has been proposed. This algorithm effectively determines the VM faults and often migrate the task to another working VM, in some cases it requires to migrate the entire batch of task that constitute the job.

## 2. Literature Survey

A redundant VM placement optimization approach in an attempt to address huge network resource consumption issue during failure recovery in cloud services is presented in [20]. The approach employs three algorithms. The first algorithm is to select a set of VM-hosting servers from a huge list of servers based on their network performance. The second is an optimal strategy which places the VMs and backups in a way to ensure k-fault-tolerance. The last one is a heuristic to address the task-to-VM reassignment optimization problem, which is formulated as finding a maximum weight matching in bipartite graphs. The spot instances as a less expensive but mostly unreliable option allow the users to bid on unused cloud computing capacity as instances and use it till the bid exceeds the current spot price. However its unreliable nature makes it more prone to faults and can cause serious delay in task completion. Therefore in an attempt to reduce the delay [12] proposes a SLA (Service Level Agreement) price history based checkpointing scheme. It aims to reduce the number of checkpoint attempts to improve the spot performance. Cloud computing can execute workflows to sustain business process management system. A lightweight checkpointing suitable to ensure fault tolerance in cloud computing during the execution of workflows has been proposed in [5]. It is an Adaptive Time based Coordinated Checkpointing ATCCp, method suitable for soft checkpointing which helps to minimize the storage time and improves consistency. Mobile cloud computing (MCC) usually means a heterogeneous mobile cloud offloading service is an attempt to enhance the performance of mobile devices. It involves a shared resource pool consisting of mobile ad-hoc networks, nearby cloudlets, and private/public cloud services. However it is prone to a variety of faults due to its ad-hoc nature. In order to improve the mobile cloud service reliability [10] presents a group based fault tolerant mechanism GFT-mCloud that classifies mobile devices into groups based on its processing capacity, mobility, and reliability. Different fault tolerance techniques are then applied to different groups based on the task offloading schedules. [25] presents a layered abstraction approach for developing and managing fault tolerance without bothering about the implementation details. This approach allows the users to just specify the desired fault tolerance level to make it operational. The current MapReduce Framework based rescheduling applied fault tolerance methods fails to keep track of the location of distributed data, the computation and storage overhead caused by the rescheduled failure tasks. However to overcome this problem [26] presents a replication-based mechanism which considers both task and node failure while computation. The energy consumption increases extensively in Cloud Systems. This calls for green computing sensitive task scheduling which attempts energy reduction while meeting the QoS requirements. [30] presents a DVFS-enabled Energy-efficient Workflow Task Scheduling algorithm (DEWTS). It attempt to reclaim slack time through merging underutilized servers in an attempt to reduce energy consumption. DEWTS calculates the initial scheduling order of all tasks, and obtains the whole makespan and deadline based on Heterogeneous-Earliest-Finish-Time (HEFT) algorithm.

## 3. Insights from Previous Work: Checkpoint Optimization

The previous work explored the possibility of using MD5 hash and Delay variation as a combined strategic parameters to detect node (VM) level Byzantine faults [1]. According to the concept, hypervisor or monitoring unit auto generate the simple message and sends it to VM at every state interval VM generate hash and communicate it back to monitoring unit. The monitoring unit measures the delay variation and changes in hash to health and integrity of every VMs deployed under it. However the massive achievement in the previous work is about optimizing the state interval to reduce the processing and cost overhead, which serves as the fundamental idea for the proposed scheduling and checkpointing algorithm discussed as follows.

Optimizing the fault tolerance techniques such as Checkpoint/Restart, job migration etc becomes a challenging task due to time, cost and space overhead it creates. This is vastly due to the reason, that the fault tolerance mechanism is performed indifferently for the entire virtual components for every regular interval without considering their performance, health etc. Therefore the goal is increase or decrease the state interval based on the performance and health of the VMs to limit the overheads to a moderate level even when the Cloud deployment grows in size and complexity such as with big data processing applications.

The following table 1 & figure 1 are constructed with delay variation ($Ɣ$) *with set of possibilities P {high =0, extreme =1}* since those are the delay that intimate that the observed virtual node may be transpiring to erroneous state or already erroneous. The other lower possibilities P {low, medium} which is extensively studied with previous work are not considered for optimization for better fault isolation. Moreover the $Ɣ¢ =0$ denotes the absence of $Ɣ$ value this denotes whether delay is low or normal and the Virtual Node (VN) is running problem free. However delay variation 'extreme' implies the VM is transpiring into a fatal stage, where as high indicate further monitoring is necessary for coming to a conclusion. Similarly, the Checksum ($¢$) variable takes on values from the set of *P {no error =0, error = 1}*. Moreover the state transition diagram and table were obtained for virtual node states *{fail-safe, Byzantine, fail-stop}* with respective state variables *{$S_0$, $S_1$, $S_2$} as follows.*

**Table 1.** State transition with Decisive Checksum and Simplified Delay Variation

| Present State | Next State | | | | Output |
|---|---|---|---|---|---|
| | 00 | 01 | 10 | 11 | |
| $S_0$ | $S_1$ | $S_2$ | $S_2$ | $S_2$ | 1 |
| $S_1$ | $S_1$ | $S_2$ | $S_2$ | $S_2$ | 1 |

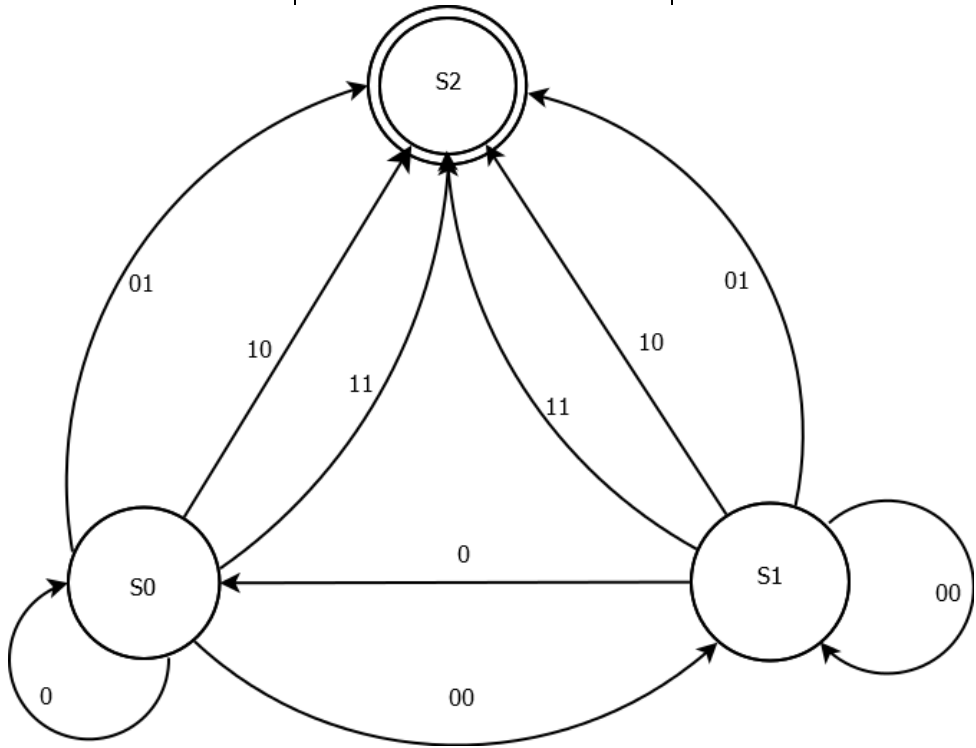

**Figure 1.** State transition with decisive Checkpoint & simplified delay variation

However if *no error* in checksum and if the delay variation is not *high* or *extreme* then the input for both $Y¢ =0$ this can happen only in state $S_1$. This implies that after exhibiting high $Y$ in previous state which caused the observing node $n_i$ to transition from $S_0 \rightarrow S_1$, it has recovered from the setback and for the current state the transition occur from $S_1 \rightarrow S_0$. Therefore in the current state observation the $Y$ seems to be either *low* or *normal* and thus makes the $Y$ input missing. This helps to generalize the nodes which though show no checksum error but shows slightly increased delay over consecutive state observations.

**Algorithm 1: State Interval Optimization ()**

```
    initialize x = j                    //where j is the pre-set initial state monitoring interval
    for each $v_i$ in $S_0$,
        if $Ɏ¢ = \{0\}$
            Assign interval j = x + j        // increasing interval
            Call Compare delay variation ()
        end if
    end for
    initialize q = 0  //to monitor staying nodes for each intervals
    for each $v_i$ in $S_1$; $Ɏ¢ == \{00\}$; q+1
            Set j = x
            Call Compare delay variation ()
            Call Checksum Challenge ()
        if q ==3
            Shutdown $v_i$
            Start new node as $v_i$
        endif
    end for
```

For the nodes that remain in $S_0$ the next interval for monitoring can be increased from interval *j* to *2j* and after *2j* it still remains in $S_0$ the interval can be further shifted to *3j* etc. Nodes in state $S_1$ alone needs frequent monitoring for every interval *j*, If the node in $S_1$ transitioned to state $S_0$ then the successive interval can be increased as long as it stays in $S_0$. However if it transitioned back to $S_1$ then the interval is reduced to initial value. Moreover for $i^{th}$ node if it stays at high for *three* successive intervals in state $S_1$ then it could be suspended for evaluation. This way the crucial overhead reduction has been achieved through minimization of the interruption time and through the signification reduction of the number of Checkpoints implemented not only for the erroneous case but also for normal case. Moreover restart can be promptly initialized with previously saved checkpoint with minimal overhead. This algorithm has been obtained as the result of working out various algorithms with state transition and mathematical modeling [1].

## 4. Proposed Work

The cloud computing involves dynamic allocation of resources that involves data centers that are often geographically distributed. The hypervisor or Virtual Machine Monitor (VMM) [28] is the high level monitoring unit that divides the available resource of the Server into various processing unit called Virtual Machines (VMs) or Virtual Nodes (VNs) and monitors their availability and performance. Based on the user request, single or more VMs are allocated to execute the submitted application. The advantage of using virtual machine is that they can execute the application across different operating systems, IDEs, or software environments. Usually the Virtual Infrastructure Management (VIM) module of Cloud computing performs resource pooling, managing physical and virtual resources etc [13]. However the VIM is not the commonly used terminology for all the cloud platforms, instead one way are another these higher management modules integrates themselves to the basic hypervisors [24]. Therefore in this research for simplicity cloud supervisor is used to mention higher management module that monitors VNs, allocates tasks to VNs, constitute Virtual Clusters and performs fault tolerance.

The next important element is Virtual Cluster (VC) [18]. Usually the virtual cluster is assumed as the dynamic grouping of various VMs. Though they are logically assumed to be in same virtual cluster but in reality, they may be placed across various physical clusters. However for better generalization in this work the virtual cluster is considered as a set of physical servers. Since most of the cases, a server with group of VMs is allocated to a Virtual Cluster [31]. Single VM for a virtual cluster is an idealistic case. This beforehand grouping of physical servers helps the CSP to provision VMs to virtual cluster dynamically upon user request or SLA agreement [8]. If CSPs does not have such beforehand knowledge allotting service to a user can be highly unrealistic and unreliable. Therefore for implementing a mission critical application or the application which involves big data processing requires the best physical servers to stand ready for dynamic allocation of VMs to make it reliable and credible.

### 4.1 Workload Sensitive Server Scheduling (WSSS)

The mission critical applications require the best available service from the Cloud service provider (CSP). The intention behind the WSSS algorithm is to keep track of every server provided to form the Virtual Cluster. It is a lightweight module and is incorporable to Cloud Supervisors. WSSS counts number of failed delay sensitive tasks which exceeds the SLA agreed QoS delays, and faults such as due to VM errors, communication errors etc. Then uses the count to rank the server after every state interval, the server with less fault counts precedes the list. This way WSSS can assist dynamic placement of jobs based on the performance of the server. Moreover it can help to rate the servers based on its previous performances with maintaining the performance status for previous Virtual Cluster implementations. Having such insight about past performance can help the management model to appropriately and dynamically choose the server for constituting VCs to run sensitive applications.

**Table 2.** Notations and Basic Definitions

| Term | Basic Definitions |
|---|---|
| $\grave{A}$ | Cloud executable application |
| $S$ | list of available servers {s1, s2, s3 …. Sn) |
| Job | Task Batch (Set of separately executable Tasks) |
| $Job_i$ | current task batch |
| $W$ | Failed workload or task (erroneous) |
| $Y$ | Failed delay sensitive workload |
| Count | Function to count the failed tasks |
| VN | Virtual node |
| WSS | Workload Sensitive Server ← List of efficient servers ranked in ascending order |
| $\Delta$ | Fault Tolerance State Interval |

**Algorithm 2: WSSS()**

Input: À , S
Split À = { Job $_1$, Job $_2$, Job $_3$, ........., Job $_n$ }
Split Job $_1$ = {Task $_{11}$, Task $_{12}$, ...... Task $_{1n}$ }
⋮
Split Job $_n$ = {Task $_{n1}$, Task $_{n2}$, ..... Task $_{nn}$}
  **for all** *s* in *S* **do**
    **if** $s_i$ is assigned to Job $_i$ **then**
      ***if*** $s_i$ not in *WSS* ***then***
        *WSS*= *WSS.increment*
        *WSSS* ← $s_i$
      **end if**
    **end if**
  **end for**
  **for** $s_1$ in WSS **do**
    **if** VN = *W* **then**
      *Count* ← *Count.increment*
    **else if** VN = *Y* **then**
      *Count*← *Count.increment*
    **end if**
    **end if**
  **end for**
//Similarly continue for $s_2$, $s_3$.... $s_n$
⋮
  **for** $s_n$ in WSS **do**
    **if** VN = *W* **then**
      *Count* ← *Count.increment*
    **else if** VN = *Y* **then**
      *Count*← *Count.increment*
    **end if**
    **end if**
  **end for**
**do** sort **WSSS (s, Count),**
  **for** i = 1 to n-1
    **if** $s_i$.count < $s_{i-1}$.count **then**
      swap ( WSSS[$s_{i-1}$], WSSS[$s_i$] )
    **end if**
  **end for**

Once the appropriate server is chosen for processing the job the next step is to monitor the VM performance for implementing appropriate fault tolerance mechanism. The monitoring is greatly discussed in the previous work [1]. Therefore the fault tolerance algorithm is discussed as follows.

### 4.2 Tactical Coordinated Checkpointing (TCC)

The previous work [2] discusses in details, about performing kernel level checkpointing for VM to achieve better task management, communication and performance. As part of the work, the management module has been extended to every VM in an effective way. Moreover a checkpoint proxy is dedicated to sort more stable storage memory to place the checkpoint files even to a remote location in an effective way. Moreover highly capable VM snapshot technique has also been devised in previous work for quick VM imaging and minimizing the storage space requirements. Now the challenge is to determine whether the checkpointing performed is going to be Independent or Synchronous [5].

Usually employed checkpointing is Synchronous, it involves at least checkpointing all the tasks involved in a job to maintain a global consistency [5]. In general all the VMs running concurrently as a part of executing an application were checkpointed at regular interval to eliminate the domino-effect that is usually caused by a single byzantine error at a single VM. This is therefore the most inefficient time, space and cost consuming model, still it is used in many cases to ensure the application safety and cloud credibility. Most often neglected and underestimated checkpointing is Independent checkpointing, because it is uncoordinated and imaged here and there without bothering about the state interval. Therefore in case if a byzantine error intrudes a single VM and go undetected then there is no state guarantee to move back to previous checkpointing but instead it often requires rollback to the initial stage. However if the independent checkpointing is done with the better detection capabilities then it can greatly reduces the time, space and cost requirements. Therefore this work combines both the checkpointing technique to come-out with an efficient hybrid checkpointing for better optimization termed as tactical coordinated checkpointing.

The following is the list of algorithm that is expected to be running while the application is assigned to the Cloud Clusters.

**Monitor** (Hash; Delay Variation)   \\ Detailed in [1], it challenges VNs with message M at optimized intervals to detect byzantine error. Every VN generates hash and sent back to cloud supervisor.

Checksum Challenge () \\ Performed by the Cloud Supervisor after Monitor () operation [1]. It generates own hash and compare it with hash generated by each VM for message M.

Compare delay variation () \\ Performed by the Cloud Supervisor after Monitor () operation. It compares the delay variation of every VM with SLA delay [1]

**Update** WSSS ( )     \\ it make sure the algorithm 2 is evoked and runs throughout execution

State Interval Optimization ($Y, \cent$) \\ calls algorithm 1 usually at every state interval for optimization

HPR_Checkpoint () \\ Detailed in [2], this algorithm covers the back-end checkpointing process in a effective way

**Algorithm 3: TCC ()**

**for each** Job$_i$
   {Task $_{i1}$, Task $_{i2}$, …… Task $_{in}$ } Run in respective {VN$_1$, VN$_2$, … VN$_i$}
   Initialize i = 1; Initialize N = 0
**for each** Δ && VN$_i$ ≤ VN$_n$
**for each** VN$_i$ **do**
   **Monitor** (Hash; Delay Variation)
   **Call** Compare delay variation ()
      Return (Ɣ)
   **End Call**
   **Call** Checksum Challenge ()
      Return (¢)
   **End Call**
   **Call** State Interval Optimization (Ɣ, ¢)
      Return (j)
   **End Call**
// j upgradable monitoring state interval, it returns either j=j or j=2j ,
//Where 2j denotes that the VN performs exceptionally well without delay variation or hash error
   **if** Δ <j **then**
     assign Δ = j
     Status = Confirmed Checkpointing
     **Call** HPR_Checkpoint ()
   **end if**
   **else if** Δ ≥ j **then**
     Status = Previous Checkpointing
     **Call** HPR_Checkpoint ()
     Start alternative VN from previous checkpointing
     N = N+1
    **if** N > 5
     Status = Job Migration
     Status = Complete Previous Checkpointing
     Halt Job   \\ halts the complete set of tasks
     **Call** HPR_Checkpoint ()
     Start alternative set of VNs from previous checkpointing
    **end if**
   **end if**
   Increment i = i +1
**end for**
**end for**
**end for**

The algorithm is the attempt to categorize the VN at every state interval into two groups based on their performance. Any VN at the given state interval will fall at the set of possibilities P

{Performing = 1, Not performing = 0}. The VN which are performing marginally that is causing some delay but not errors were also been categorized as not performing, so as to avoid all possibility of error in a mission critical application. For the performing VNs the state interval is incremented twice this ensures performance improvement and overhead reduction in terms of cost, space and time. In the previous work a separate space in the VN is reserved for periodic hash processing without interrupting the running tasks. Therefore there is no processing or performance overhead reflected to customers in implementing the Monitor (hash, checkpoint) algorithm.

## 5. State Transition Computation

The state transition is computed to understand and correlate the objective of the algorithm with its flow. Such as using the state transition computation, whether the proposed fault tolerance narrows down to byzantine error prone region or not has to be analyzed. Subsequently, whether the performance improves with increasing the state interval for VNs or not needs to be investigated. Moreover the checkpoint optimization attempt, which involves grouping the VNs into two categories with an objective to arrive somewhere between synchronous and independent checkpointing is met or not needs examination.

### 5.1 State Transition with Checkpointing Status

Initially the state transition has been obtained for TCC algorithm with Checkpointing Status (§). The states that are applicable for all the following state transitions involves the set of possibilities P *{fail-safe, Byzantine, fail-stop}* denoted with state variables *{$S_0$, $S_1$, $S_2$}* respectively. The Checkpointing Status is given as the set of possibilities {Null, Confirmed, Previous, Complete} denoted with binary values {00, 01, 10, 11}.

**Table 3.** State transition with delay variation

| Present State | Next State | | | | Output |
|---|---|---|---|---|---|
| | 00 | 01 | 10 | 11 | |
| $S_0$ | $S_0$ | $S_0$ | $S_1$ | $S_2$ | 0 |
| $S_1$ | $S_0$ | $S_0$ | $S_1$ | $S_2$ | 0 |

Among those the state $S_2$ is assigned as acceptor because if a transition ends in $S_2$ then the entire job running VNs are shut and a completely new set of VNs are started with the previous checkpointing. This marks the closing of states for the current set of VNs. Therefore the objective is never to transition into state $S_2$.

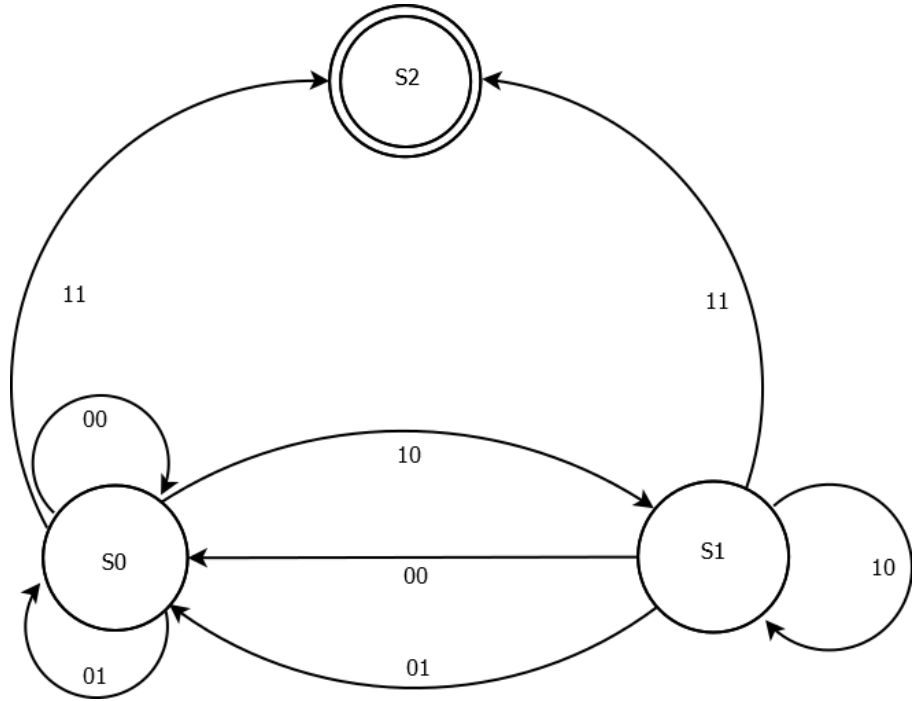

**Figure 2.** State transition diagram with Checkpointing Status

According to Figure 2 among various transitions the null transition i.e. § == '00' marks the possibility of improvisation because it denotes the set of intervals where the VN is trusted to operate without checkpointing. Initially state interval is Δ then next time it is incremented to 3Δ if there is no problem with the VNs. If the state intervals for usual case are compared with the proposed TCC case then usual states will be fixed so the series is {Δ, 2Δ, 3Δ, 4Δ …} but the interval series for TCC for error free and healthy VNs is {Δ, 3Δ, 6Δ, 12Δ …}. Hence the overhead reduction achieved with every null transition i.e for intervals {2Δ, 4Δ, 5Δ, 7Δ, 8Δ, 9Δ, 10Δ, 11Δ …} increases exponentially compare to usual case.

## 5.2 State Transition with VN Performance

The state transition for VN Performance $(Þ)$ has been developed to understand the applicability of the proposed TCC algorithm. The states *{fail-safe, Byzantine, fail-stop}* remains the same for all the cases and their respective state variables remains the same as *{$S_0$, $S_1$, $S_2$}*. In Table 2 the Input is constructed with the possibilities P {Not Performing, Performing, wary} represented by corresponding binary inputs {00, 01, 10}. Moreover the S2 remains the acceptor, in here if more VNs that is (Number >5) underperforms then only it transitions to S2 where all the VNs halt with referring to previous checkpointing such a situation is marked with a possibility 'wary' performance. However the 'Number' serves as a threshold and it can be increased if more number of VNs are running to accomplish the tasks. Say if 100 VNs is running and if ¼ of that 25 VNs experience some hiccup in delay then it indicates the necessity of the complete job migration through referring WSSS algorithm. If the running application is very critical application then 1/5 of 100 VNs that is 10 VNs experience some hiccup in delay then in this case it can considered as the necessity for complete job migration. Therefore the Threshold setting is left to the user's choice.

**Table 4.** State transition with VN Performance

| Present State | Next State | | | Output |
|---|---|---|---|---|
| | 00 | 01 | 10 | |
| $S_0$ | $S_2$ | $S_0$ | $S_2$ | 1 |
| $S_1$ | State Nullified | | | 1 |
| | | | | |

Table 4 and Figure 3 indicate that the proposed Byzantine fault tolerance operates as avoidance mechanism since it eliminates the state transition $S_0 \rightarrow S_1$. According to previous study the checkpointing is decisive in nature it detects the byzantine error with 99% accuracy most of the times but even in the worst case scenario it detects with 88% accuracy [1].

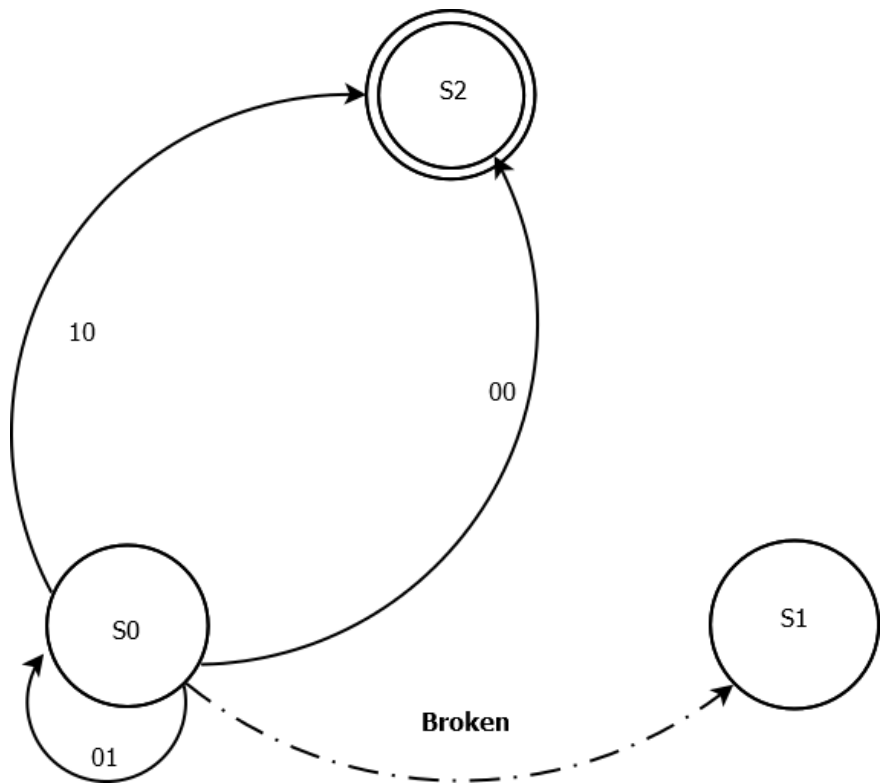

**Figure 3.** State transition diagram with checksum

This creates only 12% possibility of byzantine error getting missed out with hash based detection alone. But further analysis show that this 12% exhibits high delay variation due to the sudden inducement of foreign data in case of byzantine error. Therefore if the delay variation is 'high' then the node is checkpointed to the previous state and the node is considered not performing. This causes the checkpointing algorithm to halt the node and transfer the workload to another node from pervious checkpointing this way it effectively implements semi independent checkpointing. Moreover the 'wary' case as defined before is effective implementation of semi synchronous checkpointing. Hence according to the state transition analysis the proposed algorithm though eliminating the risk of Byzantine errors completely yet managed to optimize the fault tolerance techniques with considerable overhead reduction.

### 5.3 Byzantine Problem Tolerability Analysis

In any system a solution to the Byzantine fault tolerance usually is complex and assumed to require 3K+1 active replication system to tolerate K failures [21]. However usually the K failure in cloud system is unpredictable so the K simply be assumed as number of all running VNs i.e. K = n. This way any cloud system to run a mission critical application it requires to place K+1 replica to stand ready to replace any failed nodes. Therefore even if the k components fail it will have 1 redundant component after replicating all the failed nodes. The challenge that the Cloud faces is that, there is no certainly to say it is K-fault tolerant even after having the K+1 replacement possibilities, simply because an evasive byzantine error corrupts all the running VNs without proper detection. However, the hash and delay sensitive based detection turns out to be K+1 fault tolerant [1]. Therefore with TCC and WSSS it is modest to say it is K fault tolerant. In reality, for a mission critical application with WSSS and TCC it greatly reduces the replacement requirements to < K.

## 6. Experimental Results and Analysis

This section attempts to measure the performance of the proposed algorithms through simulating them in CloudSim [4]. CloudSim is a dedicated simulation tool for Cloud computing supports modeling the entire scenarios including data centers, VMs, VM provisioning, scheduling algorithms, fault tolerance etc.

### 6.1 System Implementation

Evaluating the proposed server level scheduler requires to run it in a massive cloud application. Therefore a huge data migration application which scales cloud clusters is programmed using CloudSim with the support of planet lab dataset. Now the voluminous data is split into various jobs and then the jobs are fragmented into workloads. The workload is bulk and requires effective server level scheduling, for that the Most Efficient Server First (MESF) [3] scheduling algorithm is implemented and compared with WSSS algorithm. MESF is chosen because it outperforms industry standard greedy algorithm [3]. Moreover it attempts to schedule the tasks to a minimum number of servers and keeps track of response time [3]. The MESF algorithm though works at server level it does not monitor the VN failures closely as WSSS algorithm. The programs for both MESF and WSSS algorithm is then developed in the CloudSim using workflowSim-01 package and tested on real-time planet lab dataset. The CloudSim version 3.03 and supporting Java version is required to configure Eclipse IDE to run both programs.

However the complete set of cloud input includes sequence of things, they are setting the parameters and implementing and calling the MESF and WSSS schedulers, initializing all the variables suitable for operating on available Planet Lab datasets. Once it is done, the relevant Scheduler that starts and allocates the VMs to the migrating workloads is run as CloudSim Program. The following is the performance comparison of the MESF and WSSS Schedulers over the same real-time migration dataset.

## 6.2 Comparing WSSS with MESF algorithm

Evaluating MESF in a homogeneous cloud environment can easily show green benefits and is capable of completing the tasks without failure [3] [10]. However evaluating the MESF and WSSS in heterogeneous environment offers insights close to real world scenario. Moreover a thorough evaluation not only considers green benefits but also considers all the metrics from time, space and cost perspectives. These are the factors usually accompany time-sensitive, delay-sensitive and space consuming workloads. Though the Cloud Services is highly dynamic in nature provisioning such sensitive and mission critical applications requires dedicated pre-allocated resources also. Therefore the limitation in handling them at scheduler level can affect the cloud service performance. The comparison result is presented in the following table.

**Table 5.** Tabulation of data over MESF algorithm

| Performance Metrics | MESF Migration | | | WSSS Migration | | |
|---|---|---|---|---|---|---|
| | | Mean | Std Dev | | Mean | Std Dev |
| Number of hosts | 800 | | | 800 | | |
| Number of VMs | 1052 | | | 1052 | | |
| Energy consumption (kWh) | 177.10 | | | 191.73 | | |
| Number of VM migrations | 23035 | | | 26634 | | |
| SLA performance degradation due to migration | 0.10% | | | 0.04% | | |
| SLA time per active host | 6.89% | | | 4.45% | | |
| Overall SLA violation | 0.12% | | | 0.07% | | |
| Average SLA violation: | 6.78% | | | 4.14% | | |
| Time before a VM migration (sec) | | 19.72 | 8.10 | | 13.97 | 6.40 |
| Execution time - VM selection (sec) | | 0.03432 | 0.02673 | | 0.00892 | 0.00941 |
| Execution time - host selection (sec) | | 0.01459 | 0.00814 | | 0.00916 | 0.00665 |
| Execution time - VM reallocation (sec) | | 0.10560 | 0.05593 | | 0.08122 | 0.04261 |
| Execution time - total (sec) | | 0.32130 | 0.24726 | | 0.20802 | 0.16453 |

According to the Table 5 it is evident that the MESF algorithm consumes more processing time due to the preprocessing of resources before allotting it to workloads. This makes it less suitable for mission critical applications. However the proposed WSSS algorithm handles workload or tasks comparatively better than the MESF algorithm. The performance of cloud on sensitive tasks has been substantially improved with the WSSS and the failure in completing the workload is effectively reduced.

Moreover after initial implementation and after conducting various test-runs with CloudSim the MESF algorithm is found to be struggling to accommodate sufficient VMs for the delay sensitive migration process. Consequently the MESF is integrated with the default fallback procedure available in the workflowsim even then the MESF algorithm finds trouble in accommodating VMs. The reason is MESF algorithm operates initially to evaluate the VMs in every server before allotting it to the workloads even when the workload is dynamically increasing, this makes the Cloud Supervisor to bypass the MESF and randomly allocate to available VMs this causes the mix-up and results in poor performance. Therefore MESF is found to be more suitable for static workload and homogeneous environment. Whereas the WSSS doesn't interrupt the initial allocation but maintains a ranking of all available servers for further allocation this way it assist the Cloud Supervisor instead of looking to overtake it. Therefore it requires no fallback procedure and it operates better than MESF.

### 6.3 Determining Performance Range

Moreover in Table 5 the values presented as mean and standard deviation marks the range of possibilities. Hence to evaluate such possibilities few important metrics are branched out and presented as Normal Distribution as follows.

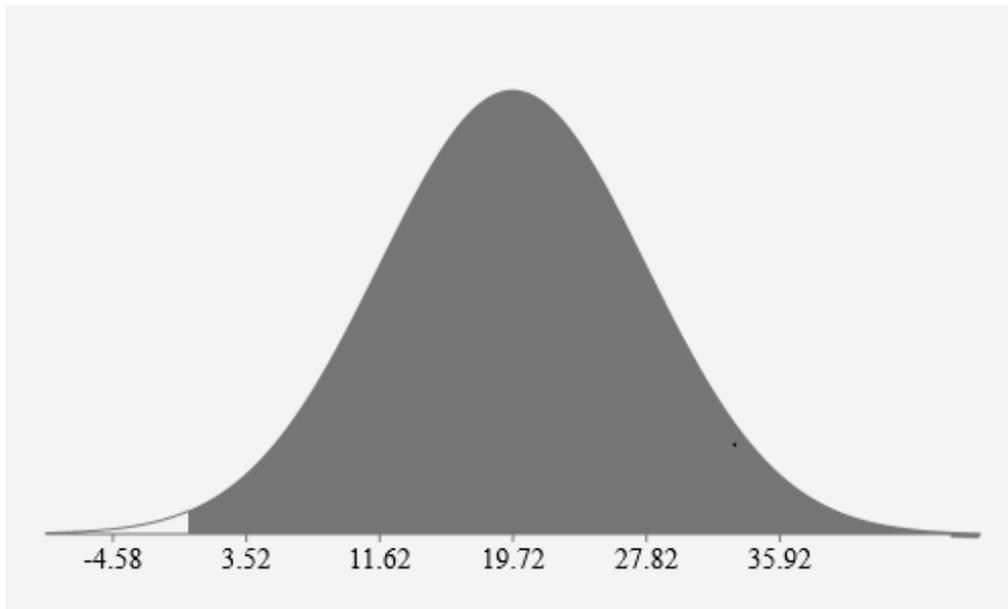

**Figure 4.** Normal Distribution of Time before a VM migration for MESF Algorithm

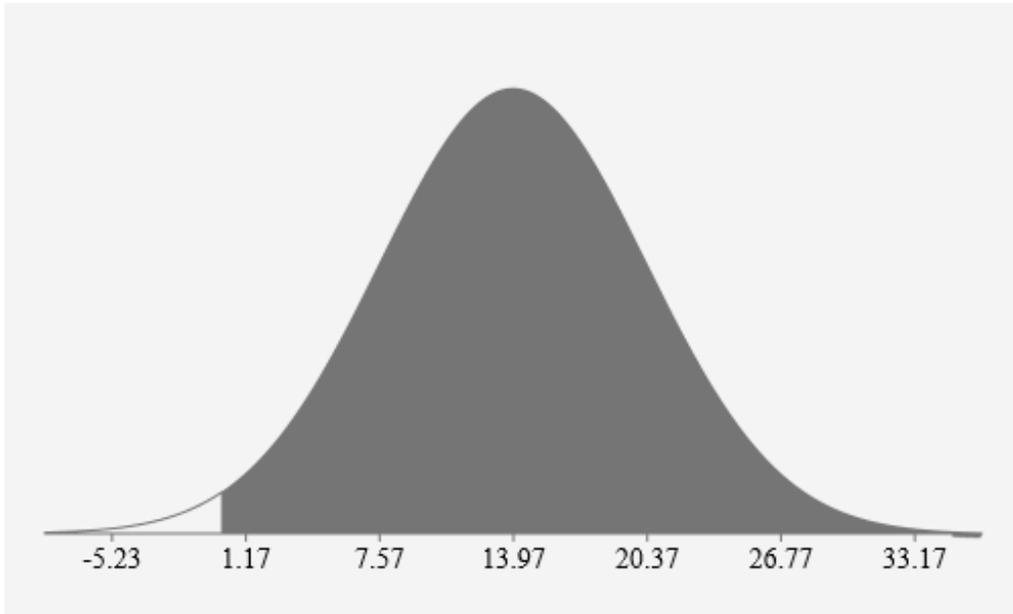

**Figure 5.** Normal Distribution of Time before a VM migration for WSSS Algorithm

The oscillation in the range of data in cloud processing can be more tightly coupled than other applications. Therefore as in Figure 4 & 5 the most occurable range of any number of instances in case of MESF is 11.62 sec to 27.82 sec where as for WSSS is 7.57 sec to 20.37 sec. In case of MESF, it takes 5 to 7 sec extra delay in allotting workloads. This can be lethal in case of delay sensitive applications.

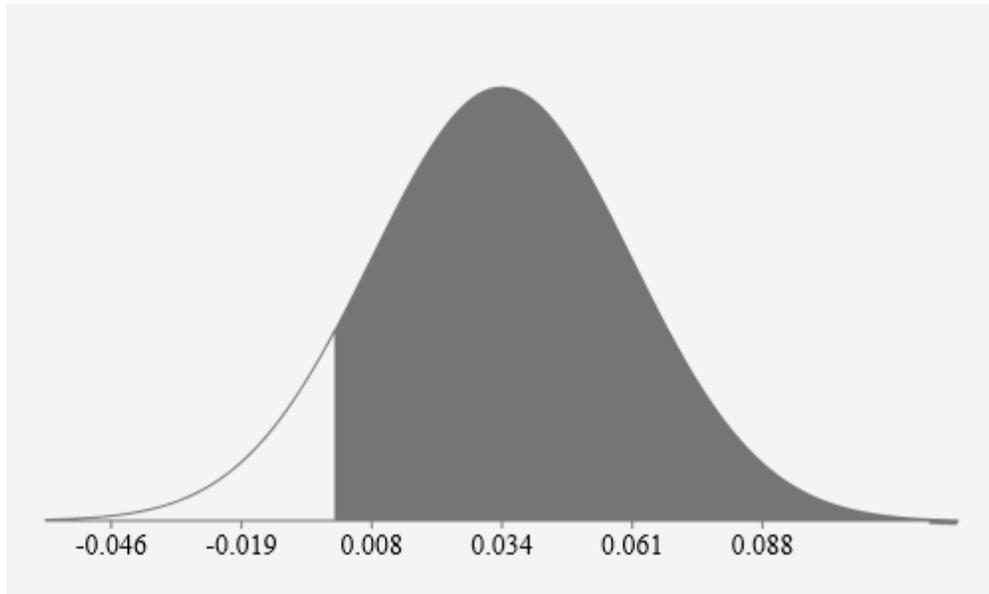

**Figure 6**. Normal Distribution for Execution time of MESF Algorithm in VM Selection

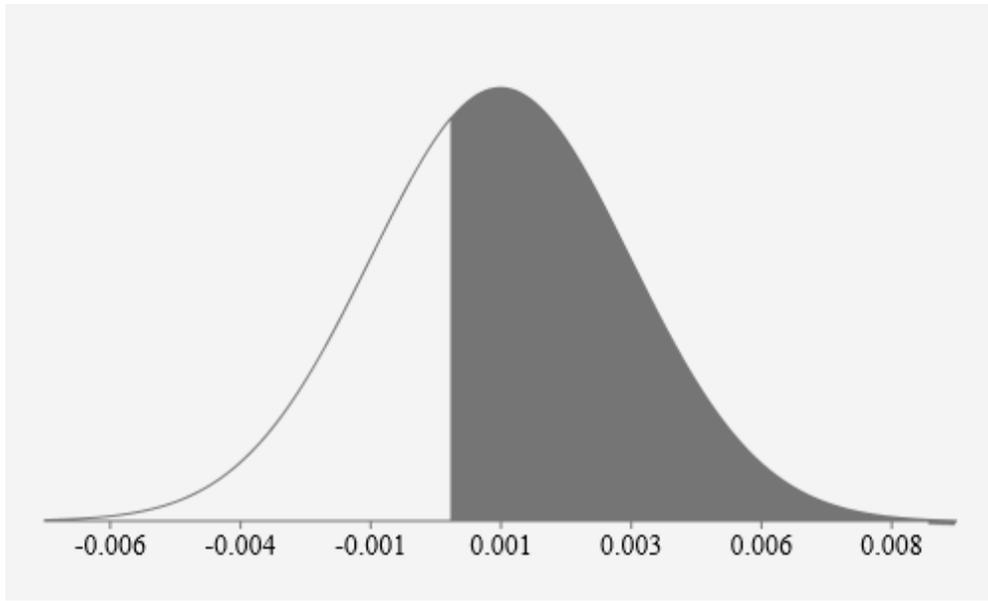

**Figure 7.** Normal Distribution for Execution time of WSSS Algorithm in VM Selection

According to the Figure 6 the MESF algorithm spends considerable time in executing algorithm before VM Selection. The probable range for such delay is mostly within the range of .008 to 0.061 seconds. In highly capable cloud computing scenario, such delay is undesirable and need to be eliminated. According to the Figure 7 the delay experienced in executing WSSS algorithm for VM Selection mostly falls within the range of 0.00l to 0.006 seconds. This is the huge improvement from MESF algorithm. However the Delay experienced is almost negligible and offers industry standard allocation condition for even the complicated Cloud migration application.

This variation of delay in Execution time of Algorithm in VM Selection and in delay before VM migration is inversely proportional to the number of workloads migrated. The simple fact that WSSS has migrated 26634 workloads, whereas MESF has migrated only 23035, itself shows that the MESF fails 3599 workloads. The range determination and analysis for two sample parameter itself proves that the MESF algorithm though tries to enhance the green computing benefits it fails to perform as WSSS algorithm. However WSSS algorithm performs better than MESF algorithm and has been proved as the practically feasible solution for real-time applications.

**Conclusion and Future Work**

A Byzantine fault often gets induced into a virtual node to yield incorrect outputs, when shared with other nodes the error can corrupt the collective outputs. Usually singe VN is made faulty to generate and propagate Byzantine errors to other VNs in quick succession. The effective detection proposed in previous work which includes hash and delay variation is used to detect Byzantine error on the fly. In this paper, a Tactically Coordinated Checkpointing (TCC) algorithm is proposed to achieve cost, space and time overhead reduction through increasing the state interval for every well performing and error free VNs. Moreover it categories even the VNs which exhibits slight increase in delay as non-performing to eliminate all the possibility of

Byzantine errors. Because previous analysis shows that the slight increase in delay is a good marker of byzantine error getting induced. Moreover for such cases, a new VN is promptly activated with previously saved error-free checkpoint. This way the proposed algorithm confines to a narrow region for eliminating the Byzantine risk completely. The state transition diagram is computed for TCC with Checkpointing Status and VN performance. The result shows, that the TCC algorithm achieves exponential overload reduction and eliminates the byzantine risks completely.

Moreover a Workload Sensitive Server Scheduling (WSSS) algorithm has been devised to identify the virtual components that are part of the Virtual Cluster in accordance to their server. Then it monitors the performance of the VNs to rate the server. This algorithm presents a counter which keeps track of the VM failures and SLA delay exceeds for every participating servers. Then based on that counter the servers are ranked in ascending order. The benefit of using WSSS is that, during execution of an application it presents the best performing servers to start backup VNs in case of failures. It also offers the history of the previous performance for making a better initial choice for running a Cloud application. It is a lightweight module and is incorporable to Cloud Supervisors. The proposed model has been simulated using the CloudSim and the results involving various performance metrics has been tabulated. The result analysis shows, that the WSSS algorithm performs better and is suitable for real-time applications.

The future work involves combining all the devised algorithms to effectively implement fault detection and tolerance into a package. The package can be then run on a test-bed like environment to perform various experimental analyses and to identify further improvement possibilities.